\documentclass[aps,prc,twocolumn,superscriptaddress,longbibliography]{revtex4-2}

\usepackage{amsmath,amssymb,physics,bm}
\usepackage{graphicx}
\usepackage{booktabs}
\usepackage{hyperref}
\usepackage{mathrsfs}
\usepackage{relsize}
\usepackage{bm}
\newcommand{\Edens}{{\mathlarger{\varepsilon}}}

\newcommand{\Lagrange}[1]{\mathscr{L}_{\raisebox{-0.5pt}{\scriptsize{#1}}}}

\begin{document}

\title{Thermodynamic versus Dynamical Description of the Neutron-Star \\
Crust-Core Instability: Implications for Crustal Observables}
\author{Athul Kunjipurayil and J. Piekarewicz}
\affiliation{Department of Physics, Florida State University, Tallahassee, FL 32306, USA}

\date{\today}
\begin{abstract}
 We investigate the crust-core transition in neutron stars using both thermodynamic and 
 dynamical descriptions of the instability. In the thermodynamic approach, the transition 
 is identified through the vanishing of a generalized incompressibility coefficient signaling 
 the onset of a bulk spinodal instability. In contrast, the dynamical approach based on the 
 relativistic random-phase approximation (RPA) incorporates Coulomb screening and 
 finite-size effects that determine the instability at finite wavelength. Using a family of 
 covariant energy density functionals spanning a broad range of symmetry-energy slopes,  
 we show that the dynamical treatment systematically predicts lower transition densities and 
 pressures compared to the thermodynamic approach. We further demonstrate that the RPA 
 instability develops at a characteristic length scale set by the competition among bulk,  
 Coulomb, and surface effects. Most importantly, we show that these differences propagate 
 directly into neutron-star observables. Because the thermodynamic approach predicts larger 
 transition pressures, it generates thicker crusts and significantly larger crustal fractions of the 
 stellar moment of inertia than the dynamical-RPA framework---with important implications 
 for the interpretation of pulsar glitches and other crust-sensitive neutron-star observables.
\end{abstract}

\maketitle

\section{Introduction}
\label{Sec:Introduction}

Neutron stars are unique laboratories for the study of neutron-rich matter under extreme conditions. 
Given their enormous density range, neutron stars are stratified into a low-density nonuniform crust 
and a high-density homogeneous core, both embedded in a neutralizing leptonic (electrons+muons)
Fermi gas\,\cite{Page:2006ud,Page:2009fu}. This remarkable stratification provides a unique opportunity 
to address one of the central questions in modern nuclear science: ``How does matter organize itself?'' 
From the formation and organization of finite nuclei in the stellar crust to the emergence of uniform matter 
in the core, neutron stars probe some of the most complex many-body dynamics that shape the visible 
universe\,\cite{LRP2015,LRP2023}.

The neutron star crust extends from densities of about $10^{4}\,{\rm g/cm^3}$ up to slightly below
nuclear-matter saturation density, $\rho_0\!\sim\!2.5\!\!\times\!10^{14}\,{\rm g/cm^3}$\,\cite{Baym:1971pw,
Ravenhall:1983uh,Lorenz:1992zz,Douchin:2000kx,Douchin:2001sv,Magierski:2001ud,Ruester:2005fm,
Oyamatsu:2006vd,Steiner:2007rr,RocaMaza:2008ja,Chamel:2008ca,Xu:2009vi,Ducoin:2010as,
Moustakidis:2010zx,Pearson:2011zz,Bertulani:2012,Utama:2015hva}. Commonly divided into an outer 
and an inner crust, matter at these sub-saturation densities organizes into a Coulomb lattice of neutron-rich 
nuclei immersed in a relativistic electron gas. As the density increases, electron capture drives the system 
toward progressively more neutron-rich nuclei until the neutron-drip line is reached at about 
$\rho\!\!\sim\!\!10^{11}\,{\rm g/cm^3}$. Beyond this threshold, nuclei can no longer bind all excess neutrons, 
and the system evolves into a crystalline lattice of neutron-rich nuclei in equilibrium with a dilute neutron 
vapor. 

The composition and structure of the crust are governed by the competition between the electronic contribution 
to the energy---which grows rapidly with density---and the nuclear symmetry energy, which penalizes large 
neutron--proton asymmetries. In particular, the density dependence of the symmetry energy plays a critical 
role in determining the proton fraction, the neutron excess of the clusters, and ultimately the location of the 
transition from the nonuniform crust to the uniform liquid core. This crust--core transition is predicted to occur 
at densities of roughly one-third to one-half of nuclear saturation density, although significant model dependence 
remains. Near the bottom of the inner crust, the characteristic length scales associated with the short-range 
nuclear attraction and the long-range Coulomb repulsion become comparable. This competition gives rise to
Coulomb frustration and the emergence of complex nonuniform structures collectively known as ``nuclear 
pasta". These exotic phases may strongly influence transport, thermal, and magnetic properties of the crust, 
as well as astrophysical observables such as neutron-star cooling and pulsar glitches\,\cite{Watanabe:2004tr,
Horowitz:2004yf,Watanabe:2005qt,Horowitz:2005zb,Maruyama:2005vb,Avancini:2008zz,Avancini:2008kg,
Newton:2009zz,Grygorov:2010zz,Schneider:2013dwa,Horowitz:2014xca,Caplan:2014gaa,Schuetrumpf:2015nza,
Fattoyev:2017zhb,Kycia:2017ibr,Nandi:2017aqq,Schneider:2018yby,Lin:2020nxy,Newton:2021vyd}.

From a theoretical perspective, the transition from the nonuniform crust to the uniform core may be viewed as 
an instability of homogeneous nuclear matter against density fluctuations. Two broad approaches are commonly 
employed to study this crust--core instability. In thermodynamic approaches\,\cite{Baym:1971pw,Ravenhall:1983uh,
Lorenz:1992zz,Kubis:2006kb}, the crust--core transition is identified through the loss of bulk stability of the uniform 
phase, typically signaled by a negative curvature of the free-energy surface as the system enters the liquid--gas 
spinodal region. In contrast, dynamical approaches---most notably those based on the relativistic random-phase 
approximation (RPA)---analyze the stability of uniform matter against finite-wavelength density fluctuations, thereby 
incorporating Coulomb, surface, and screening effects explicitly\,\cite{Baym:1971pw,Ravenhall:1983uh,Lorenz:1992zz,
Horowitz:2000xj,Carriere:2002bx}. Within the RPA framework, the crust-core transition is identified by the onset of an 
instability in the longitudinal dielectric function at finite momentum transfer, reflecting the competition among bulk, 
Coulomb, and finite-size effects.

Although both approaches aim to characterize the same physical transition, they often predict different crust-core transition 
densities, particularly for models with a stiff symmetry energy. Understanding the origin of these differences and their connection 
to the isovector sector of the nuclear interaction is one of the central goals of the present work. Equally important is assessing 
how these differences propagate to observable neutron-star properties, including the thickness and mass of the crust, the crustal 
fraction of the moment of inertia, and other quantities sensitive to the location of the crust--core interface.

The remainder of the paper is organized as follows. In Sec.\ref{Sec:Formalism} we present the thermodynamic and dynamical 
approaches to the crust--core transition, emphasizing their underlying physical differences and how such differences propagate 
into neutron-star observables. In Sec.\ref{Sec:Results} we compare the predictions of the two approaches for the transition 
properties and explore their impact on the structure of neutron stars. Finally, we summarize our findings and present our conclusions 
in Sec.\ref{Sec:Conclusions}.

\section{Formalism}
\label{Sec:Formalism}

To compute the crust-core transition density, the uniform ground state will be generated within the framework of covariant density 
functional theory (DFT). For the relativistic mean-field models employed here, the interacting Lagrangian density may be written 
as
\begin{equation}
\Lagrange{} = \Lagrange{0} + \Lagrange{1} + \Lagrange{2},
\end{equation}
where $\Lagrange{0}$  contains the free nucleon and meson contributions, while $\Lagrange{1}$ describes the 
Yukawa couplings between nucleons and the isoscalar-scalar ($\sigma$), isoscalar-vector ($\omega$), isovector-vector 
($\rho$), and photon fields\,\cite{Walecka:1974qa,Serot:1984ey,Serot:1997xg},
\begin{equation}
\!\!\!\Lagrange{1}\!=\!
\bar\psi \left[g_{\rm s}\phi   \!-\! 
    \left(g_{\rm v}V_\mu  \!+\!
    \frac{g_{\rho}}{2}{\mbox{\boldmath $\tau$}}\cdot{\bf b}_{\mu} \!+\!    
    \frac{e}{2}(1\!+\!\tau_{3})A_{\mu}\right)\!\gamma^{\,\mu}
         \right]\psi.
 \label{L1}
\end{equation}
The nonlinear meson interaction terms collected in $\Lagrange{2}$ effectively encode density-dependent effects beyond 
the original Walecka model\,\cite{Walecka:1974qa}. They are given by\,\cite{Boguta:1977xi,Serot:1984ey,
Mueller:1996pm,Lalazissis:1996rd,Serot:1997xg,Horowitz:2000xj,Todd-Rutel:2005fa,Chen:2014sca,
Chen:2014mza,Salinas:2023qic}
\begin{align}
\Lagrange{2} =    & - \frac{1}{3!} \kappa\,\Phi^3 - \frac{1}{4!} \lambda\Phi^4 
                             + \frac{1}{4!} \zeta (W_\mu W^\mu)^2 \nonumber \\
                            &   + \Lambda_{\rm v} (\boldsymbol{B}_\mu \cdot \boldsymbol{B}^{\,\mu})(W_\mu W^\mu),
\end{align}
where $\Phi\!\equiv\!g_{\rm s}\phi$, $W_\mu\!\equiv\!g_{\rm v}V_\mu$, and 
$\boldsymbol{B}_\mu\!\equiv\!g_{\rho}\boldsymbol{b}_\mu$. Additional details on how these nonlinear couplings 
dictate the density dependence of the symmetry energy may be found in Ref.\cite{Salinas:2023qic}. Having 
established the underlying covariant framework, we now turn to examine the crust-core instability.

\subsection{Crust-Core Transition: Thermodynamic Instability}
\label{sec:thermodynamic}

The stability of homogeneous neutron-star matter against coupled density and charge fluctuations was analyzed 
by Kubis in Ref.\,\cite{Kubis:2006kb} through the loss of convexity of the generalized incompressibility coefficient 
$K_{\mu}$. A concise derivation highlighting the essential steps has been relegated to an appendix. Here we establish 
a direct connection between $K_\mu$ and the determinant of the Hessian matrix governing coupled density-composition 
fluctuations.

To establish the connection between the generalized incompressibility $K_\mu$ and the Hessian matrix of second 
derivatives, we introduce the zero-temperature, constrained energy functional $\mathcal{F}$ involving both nucleonic 
($N$) and leptonic ($L$) energy density contributions:
\begin{align}
  \mathcal{F}(n,x,n_{e},n_{\mu};\mu) 
  & =  \Edens_{N}(n,x) + \Edens_{L}(n_{e},n_{\mu}) \nonumber \\
  & + \mu\left(nx\!-\!n_{e}\!-\!n_{\mu}\right) 
 \label{ConstrainedF}
\end{align}

where $n$ is the baryon density, $x$ the proton fraction, and $n_{e}$ and $n_{\mu}$ the corresponding electron
and muon densities. Here $\mu\!\equiv\!\mu_{Q}$ plays the role of the charge chemical potential conjugate to the 
conserved charge density, namely, $n_{Q}=\!n_{p}-n_{e}-n_{\mu}\!=\!0$. Note that the conservation of baryon number 
has been already enforced by writing both the neutron and proton densities in terms of the total baryon density and 
the proton fraction, namely,
\begin{equation}
 n_{n} \equiv n(1\!-\!x) \hspace{4pt}{\rm and}\hspace{4pt} n_{p} \equiv nx.
 \label{BaryonDensity}
\end{equation}
Throughout this work, leptons are treated as fully degenerate relativistic Fermi gases. 

The leptonic compositions in chemical equilibrium may be obtained after minimizing the constrained functional 
$\mathcal{F}$ with respect to the leptonic densities:
\begin{subequations}
\begin{align}
  & \left(\frac{\partial\mathcal{F}}{\partial n_{e}}\right) =0 \rightarrow \mu_{e}=\mu \rightarrow n_{e}=n_{e}(\mu), \\[0.6em]
  & \left(\frac{\partial\mathcal{F}}{\partial n_{\mu}}\right) =0 \rightarrow \mu_{\mu}=\mu \rightarrow n_{\mu}=n_{\mu}(\mu).
  \label{ChemPots}
 \end{align} 
\end{subequations}
The first relation enforces chemical equilibrium among the charged leptons and nucleons, whereas the second shows that 
the leptonic densities may be eliminated in favor of the externally fixed chemical potential $\mu\!=\!\mu_{e}\!=\!\mu_{\mu}$. 
Substituting the equilibrium leptonic densities back into Eq.(\ref{ConstrainedF}) generates the following (partially) 
Legendre-transform functional:
\begin{align}
 \mathcal{F}_{\!\mu}(n,x)  = \Edens_{N}(n,x) & + \mu nx   + \mathcal{E}_{L}(\mu) \\
                                                            = n\,{\cal U}^{N}(n,x) & + \mu nx + \mathcal{E}_{L}(\mu),
 \label{ReducedFunctional}
\end{align}
where ${\cal U}^{N}(n,x)$ is the nuclear energy per baryon and 
\begin{equation}
  \mathcal{E}_{L}(\mu) \equiv \Edens_{L}(n_{e},n_{\mu}) - \mu(n_{e}\!+\!n_{\mu}).
\end{equation}
After performing the Legendre transform, the leptonic contribution now depends only on the externally fixed charge 
chemical potential $\mu$. Thus, for fluctuations at fixed $\mu$---as in the case of Kubis\,\cite{Kubis:2006kb}---the 
curvature in the remaining $(n,x)$ subspace is governed entirely by the nucleonic part of the functional. The effect 
of the leptons is now encoded in the choice of thermodynamic ensemble through a Legendre transform between 
conjugate thermodynamic variables.

The nucleonic composition may be similarly obtained by minimizing the functional $\mathcal{F}_{\mu}$ with respect 
to the proton fraction $x$ at constant baryon density $n$. That is, the stationarity condition with respect to the proton 
fraction yields
\begin{equation}
  \left(\frac{\partial\mathcal{F}_{\!\mu}}{\partial x}\right)_{\!\!n} = 
  \left(\frac{\partial{\mathlarger{\mathlarger{\varepsilon}}}_{\!N}}{\partial x}\right)_{\!\!n} +\mu n \equiv
   n\left({\cal U}_{x}^{N}+\mu\right)=0,
  \label{muN} 
\end{equation}
where we have adopted the same notation as in Ref.\,\cite{Kubis:2006kb} by defining ${\cal U}_{x}^{N}$ as the partial derivative 
of the nuclear energy per baryon with respect to the proton fraction. This condition yields the familiar beta-equilibrium relation:
\begin{equation}
  \mu=\mu_n-\mu_p \rightarrow \mu=\mu_n-\mu_p=\mu_e=\mu_\mu.
  \label{muNL}
\end{equation}
The beta-equilibrium condition is also a direct consequence of weak-interaction equilibrium, whereby neutron beta decay and 
electron capture, 
\begin{subequations}
\begin{align}
  & n  \rightarrow p + e^- + \bar{\nu}_e, \\
  & p + e^-  \rightarrow n + \nu_e,
\end{align}
\end{subequations}
occur at equal rates in cold catalyzed neutron-star matter. Since neutrinos freely escape from the neutron star 
after the initial formation phase, their chemical potential vanishes, thereby leading directly to the beta-equilibrium 
condition given in Eq.(\ref{muNL}). Note that once the electron chemical potential exceeds the muon rest mass, 
muons appear through weak-interaction processes, so chemical equilibrium further requires $\mu_e\!=\!\mu_\mu$,
as shown in Eq.(\ref{ChemPots}).

The thermodynamic potential introduced in this section is a constrained functional in which the equilibrium configuration 
is obtained by minimizing the functional subject to the conservation of both baryon number and electric charge. Accordingly, 
fluctuations in the variables $(n,x)$ naturally describe coupled compressional and compositional modes of neutron-star matter. 

To investigate the stability of the equilibrium configuration, we examine the linear response of the system under combined
compressional and compositional modes---at fixed chemical potential---using the transformed functional $\mathcal{F}_{\mu}(n,x)$. 
It has been shown in the appendix that the Hessian matrix of second derivatives associated with $\mathcal{F}_{\mu}$ is given 
in terms of first and second derivatives of the nucleonic energy per baryon. That is,
\begin{equation}
 \hat{H}_{\mu} = \begin{pmatrix}
 \displaystyle 2\mkern2mu{\cal U}_{n}^{N}\!+\!n\,{\cal U}_{nn}^{N} & \displaystyle n\,{\cal U}_{nx}^{N}
 \\[0.6em]
\displaystyle n\,{\cal U}_{nx}^{N} & \displaystyle n\,{\cal U}_{xx}^{N}
\end{pmatrix}.
\label{HessianMatrix}
\end{equation}

The stability of the uniform phase demands that the Hessian matrix be positive definite. If instead the smaller of the two
eigenvalues vanishes, the uniform phase becomes unstable against coupled density-composition fluctuations, signaling 
the onset of the spinodal instability. Because the Hessian matrix is symmetric, namely, its eigenvalues are real and its 
determinant invariant under orthogonal transformations, the onset of the instability occurs when the smaller eigenvalue 
vanishes, or equivalently when the determinant of the Hessian matrix becomes equal to zero. That is,
\begin{equation}
 \det \hat{H}_{\mu}  
     = \left(2\mkern2mu{\cal U}_n^{N}+n\,{\cal U}_{nn}^{N}\right)\!\left(n\,{\cal U}_{xx}^{N}\right) 
      - \left(n\,{\cal U}_{nx}^{N}\right)^2 = 0.
 \label{HDeterminant}
\end{equation}
Assuming that ${\cal U}_{xx}^{N}\!>\!0$, one obtains 
\begin{equation}
 K_\mu = \frac{\det \hat{H}_{\mu}}{{\cal U}_{xx}^{N}} = 
               \left[2n\mkern2mu {\cal U}_{n}^{N}+n^{2}\,{\cal U}_{nn}^{N} 
            - \frac{\left(n\,{\cal U}_{nx}^{N}\right)^2}{{\cal U}_{xx}^{N}}\right],
 \label{KmuHess}           
\end{equation}
which is precisely the generalized incompressibility coefficient introduced by Kubis\,\cite{Kubis:2006kb} (see
also Appendix \ref{AppendixB}). The crust--core instability sets in when $K_\mu\!=\!0$.

\subsection{Crust-Core Transition: \\ Dynamic Instability}
\label{sec:dynamic}

In this section we summarize the main steps involved in computing the crust--core transition using a dynamical 
approach based on the random-phase approximation (RPA)\,\cite{Horowitz:2000xj,Carriere:2002bx}. Before 
doing so, however, it is useful to contrast the physical mechanisms underlying the transition in three commonly 
employed theoretical approaches.

In the thermodynamic approach presented in the previous section, the stability of homogeneous matter is analyzed 
in the bulk zero momentum-transfer limit ($q\!\to\!0$), where Coulomb and finite-size effects are neglected. In this 
limit, the instability is determined entirely by the curvature of the constrained functional introduced earlier and encoded 
in the Hessian matrix. The onset of the transition is signaled by the vanishing of the generalized incompressibility 
coefficient $K_\mu$, which is proportional to the determinant of the Hessian. The resulting instability therefore 
corresponds to a macroscopic spinodal decomposition of uniform matter.

Another widely used framework for describing the crust--core transition is based on the extended Thomas--Fermi (ETF) 
approximation\,\cite{Douchin:2000kx,Douchin:2001sv}. This semiclassical implementation of the local-density approximation 
evaluates the energy density locally as that of homogeneous matter, while supplementing the functional with gradient 
corrections that account for finite-size and surface effects associated with spatial density variations. Consequently, the 
energy variation generated by density perturbations contains bulk, gradient, and Coulomb contributions, leading to a 
stability matrix with an explicit momentum dependence\,\cite{Douchin:2000kx}. Stability requires the determinant of this 
matrix to remain positive for all values of $q$. As the density decreases, the determinant eventually vanishes at a characteristic 
momentum transfer, indicating that the homogeneous phase becomes unstable against finite-wavelength density fluctuations.

Finally, the relativistic random-phase approximation adopted here provides a fully microscopic linear-response treatment of 
the transition\,\cite{Horowitz:1990it,Horowitz:2000xj,Carriere:2002bx}. In contrast to the ETF approach, which relies on a 
gradient expansion of the energy functional, the RPA computes the self-consistent response of uniform neutron-star matter 
in beta equilibrium composed of neutrons, protons, and electrons. The stability of the system is probed through its static 
longitudinal response to an external perturbation carrying four-momentum $q^{\mu}\!=\!(\omega\!=\!0,\mathbf q)$. Within 
linear-response theory, the collective dynamics of the medium emerge from the interplay between polarization insertions 
generated by particle-hole excitations and the residual interaction mediated by meson and photon exchange. In relativistic 
mean-field models with nonlinear meson couplings, the propagators acquire density-dependent effective masses together 
with mixing terms between the isoscalar and isovector channels\,\cite{Carriere:2002bx}.

The stability of the homogeneous phase is determined from the longitudinal dielectric function through the condition
\begin{equation}
\epsilon_{L}(q)=\det\Big[1-D_{L}(q)\Pi_{L}(q)\Big]=0,
\label{Dielectric}
\end{equation}
where $\Pi_{L}(q)$ and $D_{L}(q)$ denote the longitudinal polarization and meson propagator matrices, respectively. 
The dielectric function emerges from iterating the particle-hole interaction to all orders, thereby modifying the analytic 
structure of the polarization insertion and promoting individual particle-hole excitations into collective modes. The onset 
of the transition is identified by the vanishing of the longitudinal dielectric function, signaling the softening of a collective 
density mode and, ultimately, the breakdown of the homogeneous phase. At the corresponding baryon density $n_c$, 
the system becomes unstable against density fluctuations characterized by a finite momentum transfer $q_c$.

Unlike the $q\!\to\!0$ thermodynamic instability discussed in Sec.\ref{sec:thermodynamic}, the RPA transition develops 
at finite momentum transfer. As in the nuclear liquid-gas transition, matter at sub-saturation density lowers its 
energy by separating into regions of density higher and lower than the average density, thereby taking advantage of the 
attractive short-range nuclear interaction. In the absence of competing effects, the system would favor the formation of 
increasingly large clusters in order to minimize the associated surface energy through a reduced surface-to-volume ratio 
and smoother density profiles. However, the long-range Coulomb interaction opposes the development of large proton-rich 
structures because concentrating electric charge over large distances becomes energetically costly. At the same time, the 
finite-range character of the nuclear force suppresses short-wavelength fluctuations, since rapidly varying density profiles 
generate sharp interfaces and large gradient energies. Consequently, surface and gradient contributions increase with 
momentum transfer, while the Coulomb interaction disfavors long-wavelength charge separation. The competition among 
these effects selects a preferred finite wavelength, $\lambda_{c}\!=\!2\pi/q_{c}$ which sets the characteristic length 
scale associated with the emerging nonuniform ``pasta" structures in the inner crust.

The dynamical transition may therefore be viewed as a finite-wavelength generalization of the thermodynamic spinodal 
instability. In the $q\!\to\!0$ limit, Coulomb and gradient contributions become negligible, and the RPA criterion reduces 
smoothly to the thermodynamic condition $K_\mu\!=\!0$. At finite momentum transfer, however, Coulomb frustration and 
surface effects shift the transition to lower densities. As we show below, RPA calculations therefore predict crust-core 
transition densities smaller than those obtained from purely thermodynamic approaches, particularly for models with a 
stiff symmetry energy.

\section{Results}
\label{Sec:Results}

\subsection{Crust-Core Transition: Characterization}

\begin{figure}[ht]
\centering
\bigskip
\includegraphics[width=\linewidth]{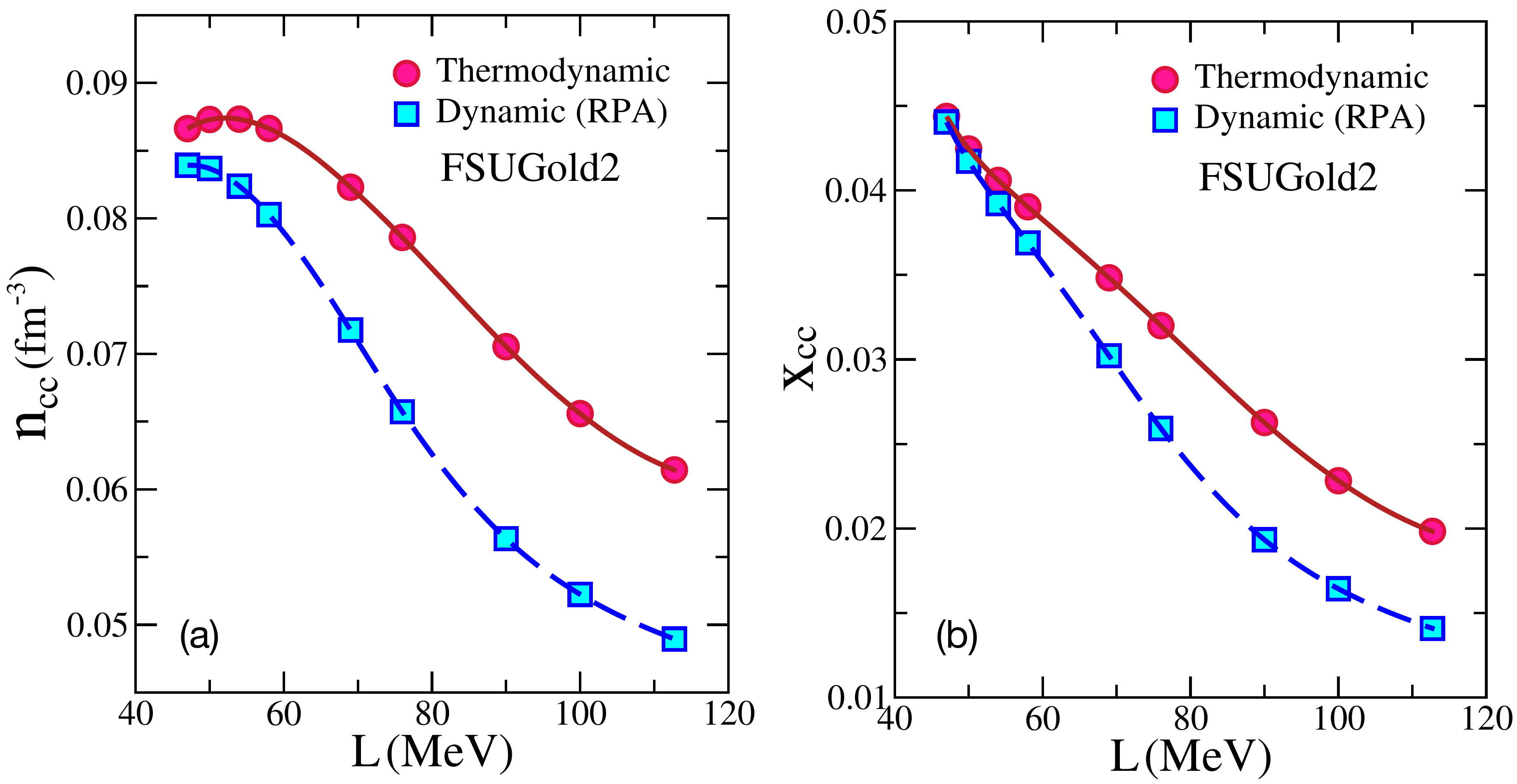}
\caption{Crust-core transition density and proton fraction as predicted by the FSUGold2 family 
	      of covariant energy density functionals\,\cite{Chen:2014sca,Reed:2021nqk}. Results
	      are displayed by both the thermodynamic and dynamical treatments of the instability.}
\label{Fig1}
\end{figure}

We start this section by displaying in Fig.\ref{Fig1} results for the crust-core transition density $n_{cc}$ and
corresponding proton fraction $x_{cc}$ as predicted by both the thermodynamic and dynamical-RPA treatments 
of the crust-core instability. Predictions are displayed by the accurately calibrated FSUGold2 covariant energy
density functional\,\cite{Chen:2014sca} alongside a set of eight systematically varied interactions with identical 
isoscalar properties as FSUGold2, but with isovector properties systematically varied in order to generate values 
for the slope of the symmetry-energy slope parameter in the range 
$L\!=\!(47\!-\!113)\,$MeV\,\cite{Chen:2014sca,Reed:2021nqk}. The figure indicates that the onset of the crust-core 
instability systematically occurs at higher transitions densities and proton fractions in the thermodynamic approach 
than in the corresponding dynamical-RPA treatment of the instability. Moreover, the discrepancy between the two 
approaches increases with $L$ suggesting that finite-size and Coulomb effects---ignored in the thermodynamic 
treatment but encoded in the momentum dependence of the RPA response---become increasingly important for 
models with stiffer symmetry energies. In the particular case of the proton fraction the observed trend may be 
attributed entirely to differences in the density dependence of the symmetry energy. As $L$ increases, the 
symmetry energy becomes softer at the sub-saturation densities relevant to the inner crust, allowing matter in beta 
equilibrium to become increasingly neutron rich. Consequently, both the thermodynamic and dynamical-RPA 
approaches predict a monotonic decrease in the transition proton fraction $x_{cc}$.

We now offer some insights into the main differences between the two approaches. Unlike the bulk 
thermodynamic instability, which corresponds to the onset of phase separation in the long-wavelength 
limit (\(q\!\to\!0\)), the dynamical instability develops at a finite momentum transfer $q_{cc}$, suggesting 
a characteristic length scale $\lambda_{cc}\!=\!2\pi/q_{cc}$. The nonzero value of $q_{cc}$ reflects the 
competition between the bulk nuclear instability that favors cluster formation and the finite-size effects 
that oppose it. This interplay may be illustrated by considering the quadratic variation of the energy 
density associated with a density fluctuation of momentum transfer $q$, which may be viewed as the 
response of uniform matter to a longitudinal probe; see Eq.(\ref{Dielectric}). In analogy with the liquid-drop 
model, this variation may be schematically decomposed into volume, surface/gradient, and Coulomb 
contributions:
\begin{align}
  \delta\mathcal{E}(q)
  & \simeq
  \frac{1}{2} \left[
    \mathcal{C}_{\rm V}(n,x)
    + \mathcal{C}_{\rm S}(n,x)\,q^{2}
    \right. \nonumber \\ 
  & \left.
    + \frac{4\pi\alpha}{q^{2}+k_{\rm TF}^{2}}\,
      \mathcal{C}_{\rm C}(n,x)
  \right]
  |\delta n(q)|^{2}.
  \label{DeltaE}
\end{align}
Here $\mathcal{C}_{\rm V}$ denotes the volume or bulk curvature that solely drives the thermodynamic 
instability, whereas $\mathcal{C}_{\rm S}q^{2}$ represents the finite-range surface or gradient penalty 
associated with momentum-dependent corrections to the short-range nuclear interaction and rapidly varying 
density profiles. The last term is the screened Coulomb contribution, with $k_{\rm TF}$ the electron 
Thomas-Fermi screening momentum. The gradient term suppresses short-wavelength fluctuations 
(large $q$), while the Coulomb interaction penalizes large-scale proton clustering and therefore suppresses 
long-wavelength charge separation (small $q$). The instability thus develops at an intermediate wavelength 
that minimizes the total energy cost. We emphasize that none of these finite-momentum contributions are 
present in the thermodynamic approach. Their inclusion in the dynamical-RPA framework delays the onset 
of the instability, shifting the crust-core transition to lower densities, as illustrated in Fig.\ref{Fig1}.

\begin{figure}[ht]
\centering
\bigskip
\includegraphics[width=\linewidth]{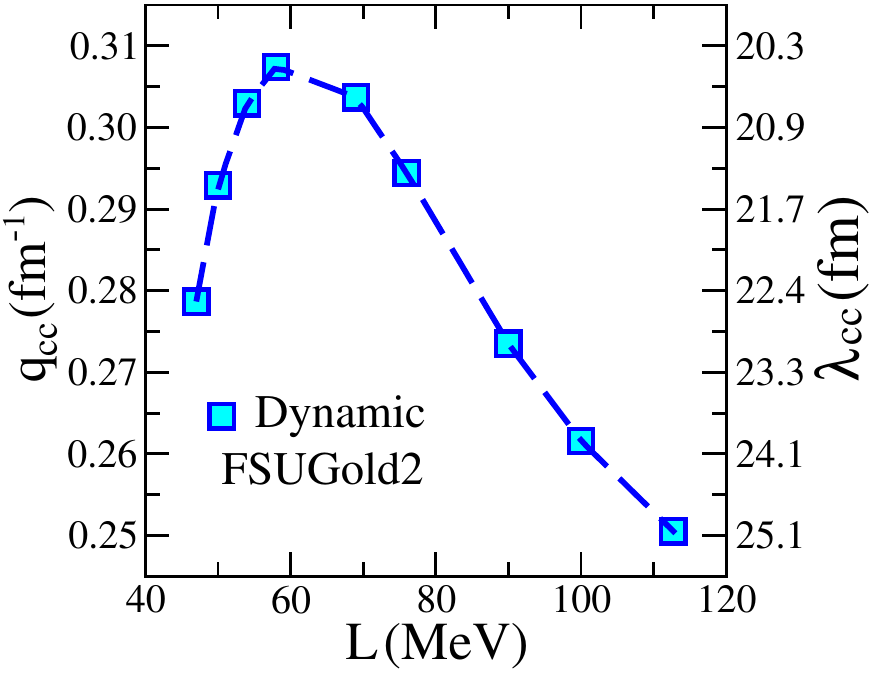}
\caption{Critical momentum transfer $q_{cc}$ associated with the onset of the dynamical-RPA 
crust-core instability as predicted by the FSUGold2 family of covariant energy density functionals. Unlike 
the thermodynamic instability, which develops in the long-wavelength limit, the dynamical instability occurs 
at finite momentum transfer, so it is characterized by a preferred wavelength $\lambda_{cc}=2\pi/q_{cc}$.}
\label{Fig2}
\end{figure}

To highlight the significant role that the slope of the symmetry energy plays in driving the RPA instability, 
we display in Fig.\ref{Fig2} the momentum transfer $q_{cc}$ at the crust-core transition for the same family
of FSUGold2 interactions shown in Fig.\ref{Fig1}. The behavior of $q_{cc}$ displayed in the figure may 
be understood qualitatively in terms of the competition between bulk, surface, and Coulomb effects. For 
models with soft symmetry energies (small $L$), both the transition density and proton fraction remain 
comparatively large. Because the electron Thomas-Fermi screening momentum scales approximately 
as $k_{\rm TF}\!\propto\!(xn)^{1/3}$, the Coulomb interaction is then strongly screened, thereby reducing 
the Coulomb penalty associated with proton clustering; see Eq.(\ref{DeltaE}). As a consequence, the 
finite-size Coulomb and gradient terms provide only modest corrections to the underlying bulk 
thermodynamic instability. Hence, the thermodynamic and dynamical-RPA transition densities remain 
close to each other, as displayed in Fig.\ref{Fig1}. As $L$ increases, matter at the transition becomes 
progressively more neutron rich, reducing the screening efficiency and enhancing the relative importance 
of finite-size effects. The instability then develops through a compromise between the surface term, which 
favors smoother density profiles and smaller values of $q$, and the screened Coulomb interaction, which 
suppresses long-wavelength charge separation and favors larger $q$. This competition initially drives a 
modest increase in $q_{cc}$. However, for sufficiently large $L$, the clusters become extremely neutron rich 
and the transition density drops significantly. Because pure neutron matter remains unbound at all densities,
the system can no longer sustain increasingly compact structures. Instead, the instability shifts toward 
smoother and more extended density fluctuations, causing the preferred momentum transfer $q_{cc}$ 
to decrease. Thus, the non-monotonic behavior of $q_{cc}$ emerges naturally from the interplay 
between the bulk thermodynamic instability and the finite-size Coulomb and surface contributions 
that ultimately fix the preferred wavelength of the unstable mode.

\begin{figure}[ht]
\centering
\bigskip
\includegraphics[width=0.8\linewidth]{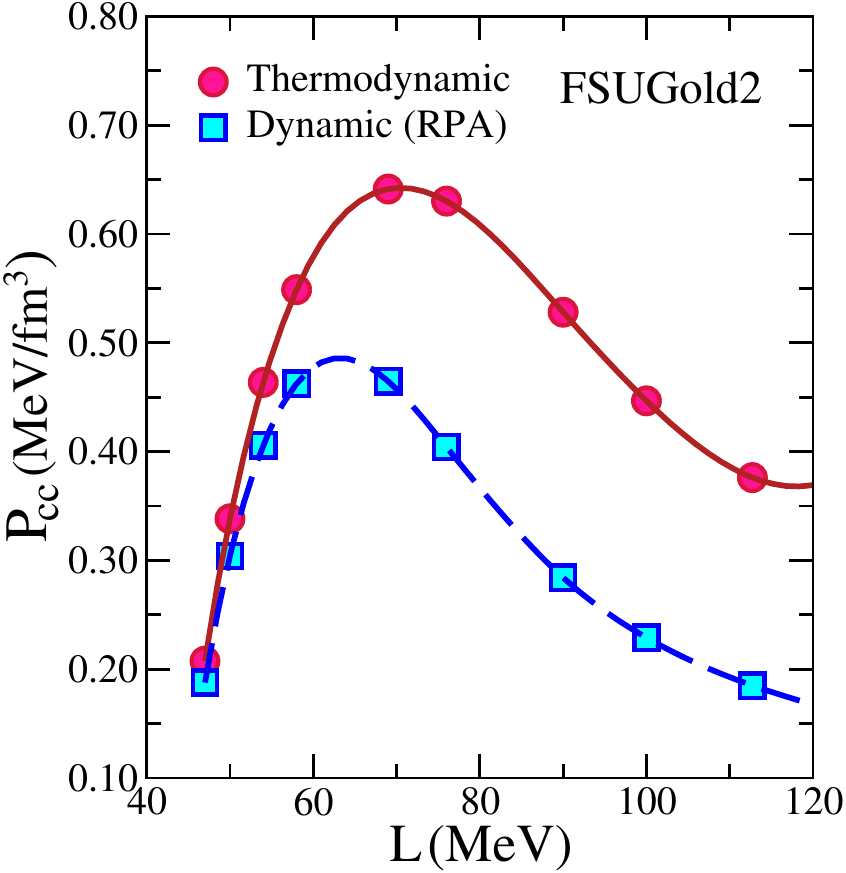}
\caption{Crust-core transition pressure as predicted by the FSUGold2 family of covariant energy 
              density functionals\,\cite{Chen:2014sca,Reed:2021nqk}. Results are displayed by both 
              the thermodynamic and dynamical treatments of the instability.}
\label{Fig3}
\end{figure}

The crust-core transition density and proton fraction displayed in Fig.\ref{Fig1} provide only part of 
the information required to characterize the transition. Equally important is the transition pressure 
$P_{cc}$, as it largely determines the thickness, mass, and moment of inertia of the crust \cite{Link:1999ca,Lattimer:2000nx,Lattimer:2006xb,Fattoyev:2010tb,Ducoin:2010as}. We therefore 
display in Fig.\ref{Fig3} the transition pressure predicted by both the thermodynamic and dynamical-RPA 
approaches for the same family of FSUGold2 interactions.

Unlike the transition density and proton fraction, both of which decrease monotonically with increasing 
$L$, the transition pressure exhibits a pronounced non-monotonic dependence on the slope of the 
symmetry energy. This behavior reflects the competing trends associated with the density dependence 
of the symmetry energy. As $L$ increases, the transition density decreases, which by itself tends to 
reduce the pressure at the crust-core interface. However, models with larger values of $L$ also predict 
a stiffer equation of state for neutron-rich matter, thereby increasing the pressure at a given density. The 
competition between these two effects generates a broad maximum in $P_{cc}$ at intermediate values 
of $L$.

Figure \ref{Fig3} also reveals a systematic difference between the two treatments of the instability. Because 
the thermodynamic approach neglects Coulomb and finite-size effects, it predicts larger transition densities 
and larger transition pressures than the dynamical-RPA framework. The discrepancy grows with increasing $L$, 
mirroring the behavior already observed for the transition density. As we demonstrate below, these differences 
have important consequences for neutron-star observables. Indeed, among all transition properties, the pressure 
at the crust-core interface provides the most direct connection between the microscopic instability and macroscopic
 crustal quantities, such as the crustal thickness and the fraction of the stellar moment of inertia residing in the crust.

\subsection{Crust-Core Transition: Implications}
Having characterized the crust-core transition in terms of its density, proton fraction, momentum transfer, and transition 
pressure, we now explore some of its astrophysical implications. Among the various observables sensitive to crustal 
properties are pulsar glitches---the sudden spin-up in the rotational frequency of a neutron star\,\cite{JBPGC,Espinoza:2011pq}. 
The glitch mechanism is intimately linked to superfluid vortices in the inner crust\,\cite{Anderson:1975zze,Alpar:1984}. 
As the pulsar spins down due to the emission of electromagnetic radiation, the vortex distribution---pinned to the crystal 
lattice of neutron-rich nuclei---falls out of equilibrium. This creates a differential rotation between the slower crust and 
the faster superfluid. When the lag becomes critical, a fraction of the vortices abruptly unpin, migrate outward, and 
transfer their angular momentum to the solid crust, triggering a glitch.

Accounting for the large, regular Vela glitches originally required only about $1.6\%$ of the stellar moment of inertia to 
reside in the crust\,\cite{Link:1999ca,Hooker:2013fda}. However, incorporating crustal entrainment increases this 
requirement to roughly $7$--$9\%$\,\cite{Chamel:2012ae,Andersson:2012iu,Piekarewicz:2014lba}. Because entrainment 
reduces the effective angular-momentum reservoir available in the crustal superfluid, early studies suggested that the 
crust alone may not provide a sufficiently large moment of inertia to account for the observed glitches\,\cite{Andersson:2012iu}. 
It was later shown, however, that uncertainties in the density dependence of the symmetry energy permit models with 
substantially larger crustal moments of inertia, thereby alleviating this apparent tension\,\cite{Piekarewicz:2014lba}. In 
this section we demonstrate that the larger transition pressures predicted by the thermodynamic approach generate 
thicker crusts and consequently larger values of $I_{\rm crust}/I$, further enhancing the ability of the crustal superfluid
to account for the observed Vela glitches.

\begin{figure}[ht]
\centering
\bigskip
\includegraphics[width=0.85\linewidth]{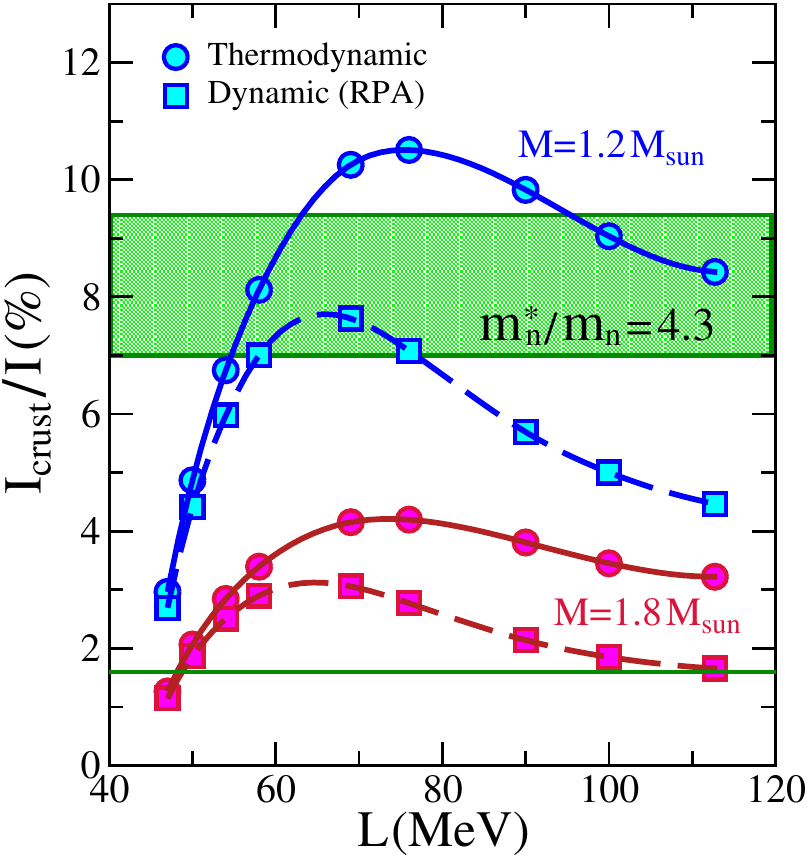}
\caption{Fraction of the stellar moment of inertia residing in the crust as a function of the symmetry-energy slope
$L$ for the FSUGold2 family of models. Results obtained using the thermodynamic and dynamical (RPA) 
crust-core transitions are shown by circles (solid lines) and squares (dashed lines), respectively. The horizontal 
line and shaded band indicate glitch constraints without and with crustal 
entrainment\,\cite{Andersson:2012iu,Chamel:2012ae}.}
\label{Fig4}
\end{figure}

We display in Fig.\ref{Fig4} the fraction of the stellar moment of inertia residing in the crust, $I_{\rm crust}/I$, for 
FSUGold2\,\cite{Chen:2014sca} and eight systematically varied interactions that share the same isoscalar properties 
but differ in the slope of the symmetry energy $L$\,\cite{Reed:2021nqk}. Results obtained using the thermodynamic 
crust-core transition are shown by circles connected with solid lines, while those based on the dynamical RPA transition 
are shown by squares connected with dashed lines. For clarity, predictions are displayed for neutron stars with masses 
of $M\!=\!1.2\,M_{\odot}$ and $M\!=\!1.8\,M_{\odot}$.

The horizontal line in the figure corresponds to the observational constraint inferred from Vela glitches under the assumption 
of negligible crustal entrainment. In this case, long-term timing observations require that at least $1.6\%$ of the stellar 
moment of inertia reside in the angular-momentum reservoir\,\cite{Link:1999ca}. The inclusion of crustal entrainment 
substantially strengthens this requirement. Indeed, adopting an effective neutron mass of 
$m_n^{\star}/m_n\!=\!4.3$\,\cite{Andersson:2012iu} raises the lower limit to nearly $7\%$, while band-structure calculations 
suggest that the required fraction could be as large as $9.4\%$\,\cite{Chamel:2012zn,Chamel:2012ae}. These constraints 
are indicated by the shaded band in the figure.

Before discussing the results, it is important to note that $I_{\rm crust}/I$ exhibits a non-monotonic dependence on $L$. 
This contrasts with both the transition density $n_{\rm cc}$ and proton fraction $x_{\rm cc}$, which vary monotonically 
with $L$ (see Fig.\ref{Fig1}). The origin of this behavior lies in the transition pressure, the quantity that primarily controls 
the crustal moment of inertia and whose dependence on $L$ is itself non-monotonic; see Refs.\cite{Link:1999ca,Lattimer:2000nx,
Lattimer:2006xb,Fattoyev:2010tb,Ducoin:2010as}. Consequently, the largest values of $I_{\rm crust}/I$ are expected for 
models with intermediate values of $L$.

Figure \ref{Fig4} confirms this expectation. Indeed, the maximum crustal fraction occurs at $L\!\approx\!75\,{\rm MeV}$ for 
the thermodynamic approach and at $L\!\approx\!65\,{\rm MeV}$ for the dynamical treatment of the instability. More 
importantly, the thermodynamic transition systematically predicts larger values of $I_{\rm crust}/I$ than the dynamical 
transition, reflecting its larger transition density and, therefore, larger crustal thickness. We observe that in the absence 
of entrainment, all models comfortably satisfy the Vela constraint for both stellar masses. Once entrainment is included, 
however, the situation becomes significantly more constrained. The figure suggests that explaining the Vela glitches with 
the crustal superfluid alone remains plausible for relatively low-mass neutron stars, but becomes increasingly difficult for 
massive stars, particularly for $M\!\approx\!1.8\,M_\odot$.

\begin{figure}[ht]
\centering
\bigskip
\includegraphics[width=0.85\linewidth]{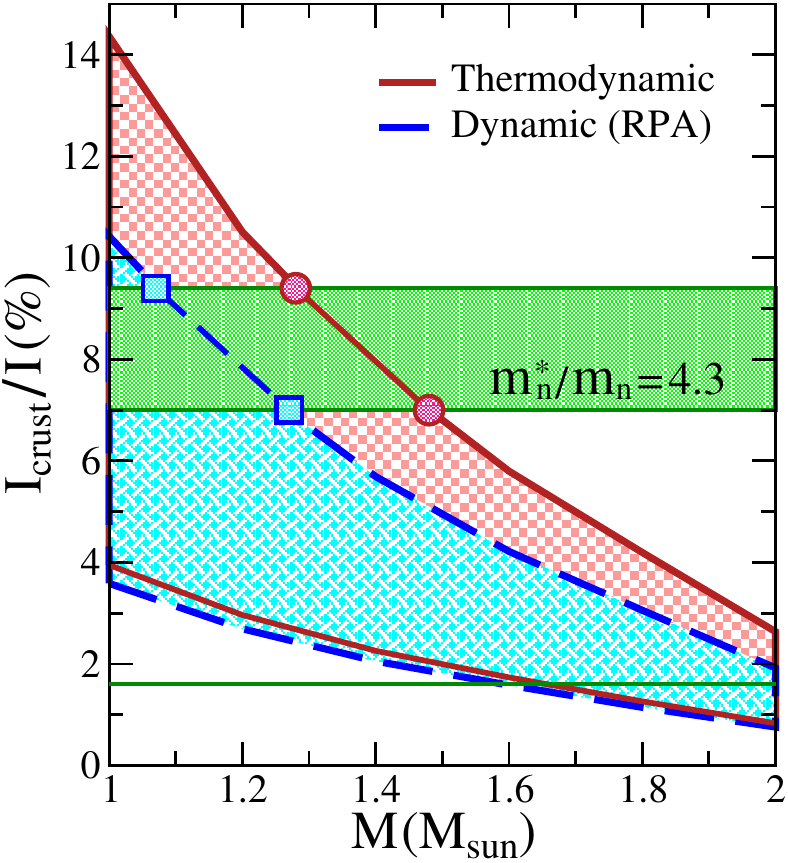}
\caption{Fraction of the stellar moment of inertia residing in the crust as a function of the stellar mass
for the FSUGold2 family of models. Results obtained using the thermodynamic and dynamical (RPA) 
crust-core transitions are shown by circles (solid lines) and squares (dashed lines), respectively. The horizontal 
line and shaded band indicate glitch constraints without and with crustal 
entrainment\,\cite{Andersson:2012iu,Chamel:2012ae}.}
\label{Fig5}
\end{figure}

To further quantify the impact of the treatment of the crust-core instability, we display in Fig.\ref{Fig5} the 
fraction of the stellar moment of inertia residing in the crust as a function of stellar mass. As in Fig.\ref{Fig4}, 
the horizontal line and shaded band represent the constraints inferred from Vela glitches without and with 
crustal entrainment, respectively.

In both the thermodynamic and dynamical-RPA approaches, the lower boundaries of the bands are defined 
by the model with the softest symmetry energy, $L\!=\!47\,{\rm MeV}$, which yields the smallest values of 
$I_{\rm crust}/I$. In contrast, the upper boundaries are generated by models with intermediate values of $L$, 
reflecting the non-monotonic dependence of the crustal moment of inertia on the symmetry-energy slope. 
Specifically, the largest crustal fractions are obtained for $L\!=\!76\,{\rm MeV}$ in the thermodynamic approach 
and for $L\!=\!69\,{\rm MeV}$ in the dynamical approach.

Although the differences between the two approaches are modest near the lower boundaries, they become 
significant near the upper boundaries, where the crustal moment of inertia is largest. Indeed, the thermodynamic 
approach systematically predicts larger values of $I_{\rm crust}/I$ than the dynamical-RPA approach over the 
entire mass range. As a consequence, the thermodynamic treatment allows the large Vela glitches to be 
explained---even in the presence of substantial crustal entrainment---for neutron stars with masses up to 
approximately $1.5\,M_{\odot}$. In contrast, the corresponding upper limit is reduced to about $1.3\,M_{\odot}$ 
when the dynamical-RPA transition is employed.

This sensitivity of the crustal moment of inertia to the treatment of the crust-core instability constitutes the 
central result of this work. Because crustal observables depend directly on the transition density and pressure, 
uncertainties in the underlying instability mechanism propagate into predictions for glitch phenomenology and 
other observables  highly sensitive to the neutron-star crust.

\section{Conclusions}
\label{Sec:Conclusions}

We have investigated the neutron-star crust-core transition using two complementary descriptions of the instability 
of homogeneous matter: a thermodynamic approach based on the generalized incompressibility coefficient $K_\mu$ 
and a dynamical approach based on the relativistic random-phase approximation. Although both approaches identify 
the same physical transition from uniform matter to the nonuniform inner crust, they differ fundamentally in their 
treatment of finite-size effects. The thermodynamic approach probes the bulk $(q\!\rightarrow\!0)$ spinodal instability, 
whereas the RPA framework incorporates Coulomb screening and finite-range nuclear interactions and therefore 
determines the onset of the instability at a finite momentum transfer $q_c$.

Using a family of systematically varied FSUGold2 energy density functionals spanning a broad range of symmetry-energy 
slopes $L$, we quantified the impact of the differences on the treatment of the crust-core instability. We found that the 
dynamical-RPA treatment systematically predicts lower transition densities and proton fractions than the thermodynamic 
approach, with the discrepancy increasing as the symmetry energy becomes stiffer. We further demonstrated that the 
instability develops at a characteristic finite wavelength determined by the competition among bulk, Coulomb, and surface 
effects. The corresponding critical momentum transfer exhibits a non-monotonic dependence on $L$, reflecting the evolving 
balance among these competing mechanisms.

The differences between the two instability criteria propagate directly into neutron-star observables through their impact 
on the transition density and, especially, the transition pressure. Because the thermodynamic approach predicts 
systematically larger transition pressures, it generates thicker crusts and larger crustal fractions of the stellar moment of 
inertia. This effect is particularly important in the context of pulsar glitches. We found that the thermodynamic treatment 
permits significantly larger values of $I_{\rm crust}/I$ than the dynamical-RPA approach, thereby extending the range of 
stellar masses for which the large glitches observed in the Vela pulsar may be explained within a crustal-superfluid scenario, 
even in the presence of substantial crustal entrainment. In particular, the maximum stellar mass compatible with the Vela 
glitch constraints increases from approximately $1.3\,M_\odot$ in the dynamical-RPA approach to about $1.5\,M_\odot$ in 
the thermodynamic treatment. This sensitivity of crustal observables to the treatment of the crust-core instability constitutes 
the central finding of the present work.

Although we have focused here on the crustal moment of inertia, the influence of the crust-core transition extends far 
beyond glitch phenomenology. The transition density and pressure determine the thickness and mass of the crust, the 
extent of the nuclear-pasta region, and the elastic, thermal, and transport properties of neutron-star matter. These 
quantities impact a broad range of observables, including magnetar quasi-periodic oscillations and crustal shear 
modes\,\cite{Samuelsson:2007ug,Sotani:2007dr,Steiner:2009hj}, the thermal evolution and relaxation of neutron 
stars\,\cite{Brown:2009,Page:2010aw}, and the electrical and thermal conductivities associated with the nuclear-pasta 
phases near the crust-core interface\,\cite{Horowitz:2014xca}. Crustal properties may also influence the tidal
 response of neutron stars and the dynamics of compact-binary inspirals\,\cite{Tsang:2012}. Future work, by us and
 others, should therefore explore how the differences between thermodynamic and dynamical instability criteria 
 propagate into these observables and whether current and forthcoming multimessenger observations can help 
 constrain the crust-core transition.

\appendix
\section{Hessian Formulation of the Thermodynamic Instability}
\label{AppendixA}

In this appendix we compute the matrix elements of the $2\!\times\!2$ Hessian matrix of second derivatives. The Hessian 
associated with the constrained functional $\mathcal{F}_{\!\mu}(n,x)$ defined in Eq.(\ref{ReducedFunctional}) is given by
\begin{equation}
\hat{H}_\mu \equiv
\begin{pmatrix}
\displaystyle\left(\frac{\partial^2\!{\cal F}_{\!\mu}}{\partial n^2}\right)
&
\displaystyle\left(\frac{\partial^2\!{\cal F}_{\!\mu}}{\partial n \partial x}\right)
\\[1.4em]
\displaystyle\left(\frac{\partial^2\!{\cal F}_\mu}{\partial x \partial n}\right)
&
\displaystyle\left(\frac{\partial^2\!{\cal F}_\mu}{\partial x^2}\right)
\end{pmatrix}.
\label{HMatrix}
\end{equation}
To compute the individual matrix elements, we first evaluate the relevant first derivatives:

\begin{subequations}
\begin{align}
  & \left(\frac{\partial{\cal F}_{\!\mu}}{\partial n}\right) = {\cal U}^{N} + n\,{\cal U}^{N}_{n} +\mu x ,  \\[0.4em]
  & \left(\frac{\partial{\cal F}_{\!\mu}}{\partial x}\right) = n\,{\cal U}^{N}_{x} + \mu n.
\end{align}
\end{subequations}
In turn, the set of second derivatives are given by
\begin{subequations}
\begin{align}
  & \left(\frac{\partial^2\!{\cal F}_{\!\mu}}{\partial n^2}\right) = 2\,{\cal U}^{N}_{n}+n\,{\cal U}^{N}_{nn}, \\
  & \left(\frac{\partial^2\!{\cal F}_{\!\mu}}{\partial n \partial x}\right) = 
        n\,{\cal U}^{N}_{nx} + {\cal U}_{x}^{N} + \mu \rightarrow n\,{\cal U}^{N}_{nx}, \\
  & \left(\frac{\partial^2\!{\cal F}_{\!\mu}}{\partial x \partial n}\right) = 
        n\,{\cal U}^{N}_{nx} + {\cal U}_{x}^{N} + \mu \rightarrow n\,{\cal U}^{N}_{nx}, \\
  & \left(\frac{\partial^2\!{\cal F}_\mu}{\partial x^2}\right) = n\,{\cal U}^{N}_{xx},
\end{align}
\end{subequations}
where in deriving the mixed derivative we used the chemical equilibrium condition ${\cal U}^{N}_{x} \!+\!\mu\!=\!0$.
These are the relevant matrix elements of the Hessian displayed in Eq.(\ref{HessianMatrix}).

\section{Derivation of the Generalized Incompressibility $K_{\mu}$}
\label{AppendixB}

In the thermodynamic framework presented by  Kubis\,\cite{Kubis:2006kb}, the stability of 
homogeneous neutron-star matter may be analyzed in terms of the convexity of the energy 
per baryon with respect to coupled density and charge fluctuations. The starting point in the
derivation is the total energy per baryon which may be written as
\begin{equation}
{\cal U}(n,x,\mu) = {\cal U}^{N}(n,x)+\frac{\mathlarger{\mathlarger{\varepsilon}}_{L}(\mu)}{n},
\label{Unxy}
\end{equation}
where 
\begin{subequations}
\begin{align}
 & {\cal U}^{N}(n,x)\!\equiv\!\frac{\Edens_{N}(n,x)}{n}, \\
 & \Edens_{L}(\mu) \equiv \Edens_{\rm e}(n_{e}) + \Edens_{\rm \mu}(n_{\mu}).
\end{align}
\end{subequations} 
Although the total energy per baryon contains both nucleonic and leptonic contributions, the fixed-$\mu$ 
thermodynamic derivatives entering $K_{\mu}$ ultimately depend only on the nucleonic sector once
 chemical equilibrium is imposed.

In turn, the total pressure is similarly decomposed into nucleonic ($N$) and leptonic ($L$) contributions.
That is,
\begin{equation}
P(n,x,\mu) = P_{N}(n,x) + P_L(\mu).
\end{equation}
Note that for the nucleonic component, we continue to use the conserved baryon density $n$ and the 
proton fraction $x$, rather than the individual neutron and proton densities. Also note that the leptonic 
energy density as well as its associated pressure are determined---after enforcing chemical 
equilibrium---solely by the chemical potential $\mu$.

Following Ref.\,\cite{Kubis:2006kb}, the relevant thermodynamic quantity controlling the onset of the 
instability is the generalized incompressibility at fixed chemical potential:
\begin{equation}
 K_\mu \equiv \left(\frac{\partial P}{\partial n}\right)_{\!\!\mu}.
 \label{Kmu0}
\end{equation}
The chemical potential $\mu$ is fixed by beta equilibrium. That is, from Eq.(\ref{muN}) we obtain
\begin{equation}
\mu = \mu_n\!-\!\mu_p = -\left(\frac{\partial\,{\cal U}^{N}}{\partial x}\right)_{\!\!n} \!\!\! \equiv -{\cal U}_x^N.
\label{BetaEq2}
\end{equation}

Because the incompressibility in Eq.(\ref{Kmu0}) must be computed at fixed chemical potential, then the
proton fraction $x$ must adjust itself to the density changes in order to maintain the chemical potential 
fixed. To determine $(\partial x/\partial n)_\mu$ we make use of Eq.(\ref{BetaEq2}). That is,
\begin{align}
  d\mu &=  \left(\frac{\partial\mu}{\partial n}\right)_{\!\!x} dn + \left(\frac{\partial\mu}{\partial x}\right)_{\!\!n} dx \nonumber\\
           & =-{\cal U}_{nx}^{N}\,dn - {\cal U}_{xx}^{N}\,dx = 0.
\label{dmu}
\end{align}
which results in 
\begin{equation}
\left(\frac{\partial x}{\partial n}\right)_\mu = - \left(\frac{{\cal U}_{nx}^N}{{\cal U}_{xx}^N}\right).
\label{dxdn}
\end{equation}

The generalized incompressibility may now be evaluated by differentiating the nucleonic pressure as follows:
\begin{equation}
K_{\mu} =
\left(\frac{\partial P}{\partial n}\right)_{\!\!\mu}  \!=\!
\left(\frac{\partial P_N}{\partial n}\right)_{\!\!x} \!+\!
\left(\frac{\partial P_N}{\partial x}\right)_{\!\!n}
\left(\frac{\partial x}{\partial n}\right)_{\!\!\mu} \\[0.4em]
\label{KmuAppB}
\end{equation}
Given that the nuclear contribution to the pressure at $T\!=\!0$ may be written as 
\begin{equation}
P_N \!=\!-\!\left(\frac{\partial E_{N}}{\partial V}\right)_{\!\!\!N} 
         \!\!\!= n^2\left(\frac{\partial\,{\cal U}^N}{\partial n}\right) = n^2\,{\cal U}_n^N,
\end{equation}
the relevant derivatives of the pressure entering Eq.(\ref{KmuAppB}) are given by
\begin{subequations}
\begin{align}
 \left(\frac{\partial P_N}{\partial n}\right)_{\!\!x} & = 2n\,{\cal U}_n^N+n^2\,{\cal U}_{nn}^N, \\
 \left(\frac{\partial P_N}{\partial x}\right)_{\!\!n} & = n^2\,{\cal U}_{nx}^N.
\end{align}
\end{subequations}
Hence, one finally obtains the generalized incompressibility at constant chemical potential as an expression given entirely by first and second derivatives of the nucleonic contribution to the energy per nucleon. That is,
\begin{equation}
 K_\mu = \left[2n\mkern2mu {\cal U}_{n}^{N}+n^{2}\,{\cal U}_{nn}^{N} 
            - \frac{\left(n\,{\cal U}_{nx}^{N}\right)^2}{{\cal U}_{xx}^{N}}\right],
\label{KmuFinal}
\end{equation}
which corresponds to Eq.\,(15) of Kubis\,\cite{Kubis:2006kb} and is proportional to the determinant of the Hessian
matrix listed in Eq.(\ref{KmuHess}). As was shown before, at $K_\mu\!=\!0$ the uniform phase becomes unstable 
against coupled density-composition fluctuations, signaling the onset of the spinodal instability. Although the final 
expression for $K_{\mu}$ depends solely on derivatives of the nuclear energy, the leptonic sector still plays a fundamental 
role. Indeed, leptons determine the chemical potential $\mu$ and enforce beta equilibrium, thereby fixing the trajectory 
$x(n)$ along which the pressure derivative is evaluated. 

\bibliographystyle{apsrev4-2}
\bibliography{Main.bbl}

\begin{acknowledgments}\vspace{-10pt}
This material is based upon work supported by the U.S. Department of Energy Office of Science, 
Office of Nuclear Physics under Award Numbers DE-FG02-92ER40750.
\end{acknowledgments}

\end{document}